\documentclass[aps,prl,twocolumn,groupedaddress,showpacs]{revtex4}
\usepackage{epsf,color,soul}

\usepackage[dvips]{epsfig}

\scrollmode
\begin{document}


\title{Controlled light-matter coupling for a single quantum dot embedded in a pillar microcavity using far-field optical lithography  }



\author{A. Dousse, L. Lanco$^{\dag}$ J. Suf\mbox{}fczy\'nski, E. Semenova, A. Miard, A. Lema\^itre, I.
Sagnes, C. Roblin, J. Bloch and P. Senellart }

\affiliation{Laboratoire de Photonique et Nanostructures, LPN/CNRS,
Route de Nozay, 91460 Marcoussis, France
\\$^{\dag}$also Universit\'e Paris Diderot, 10 rue Alice Domon et L\'eonie
Duquet, 75205 Paris, France}

\email[]{pascale.senellart@lpn.cnrs.fr}


\date{\today}

\begin{abstract}
Using far field optical lithography, a single quantum dot is
positioned within a pillar microcavity with a 50 nm accuracy. The
lithography is performed in-situ at 10 K while measuring the
quantum dot emission. Deterministic spectral and spatial matching
of the cavity-dot system is achieved in a single step process and
evidenced by the observation of strong Purcell effect.
Deterministic coupling of two quantum dots to the same optical
mode is achieved, a milestone for quantum computing.
\end{abstract}

\pacs{78.55.Cr, 78.67.Hc, 42.50 Pq., 81.16.Nd}

\maketitle


Quantum emitters such as quantum dots \cite{1}, colloidal quantum
dots \cite{colloidal} or  single fluorescent molecule
\cite{molecule} are usually randomly distributed on the surface of
a substrate and suffer from broad inhomogeneous distributions of
their spectral lines. Although  spectral identification and
spatial localization of individual quantum emitters is made
possible using high resolution confocal micro-spectroscopy, the
fabrication of devices involving dots of known position and
spectral characteristics remains a serious experimental challenge.

\noindent An example of such  device is a quantum dot (QD) which
emission and location is matched to an optical microcavity mode.
Such devices are desirable for high throughput and directional
single photon sources \cite{2,3,4,5} and for building blocks for
quantum computing \cite{6,7,8,9}. A deterministic and scalable
fabrication method allowing choosing a single QD of given spectral
characteristics to be inserted in an optical device with an
accuracy of few tens of nanometers is missing. This is a
prohibiting barrier to the development of solid state quantum
emitter devices as well as to comprehensive fundamental
investigations of solid state cavity quantum electrodynamics
(CQED).

\begin{figure}[t]
\begin{center}
\rotatebox{270}{\epsfig{file=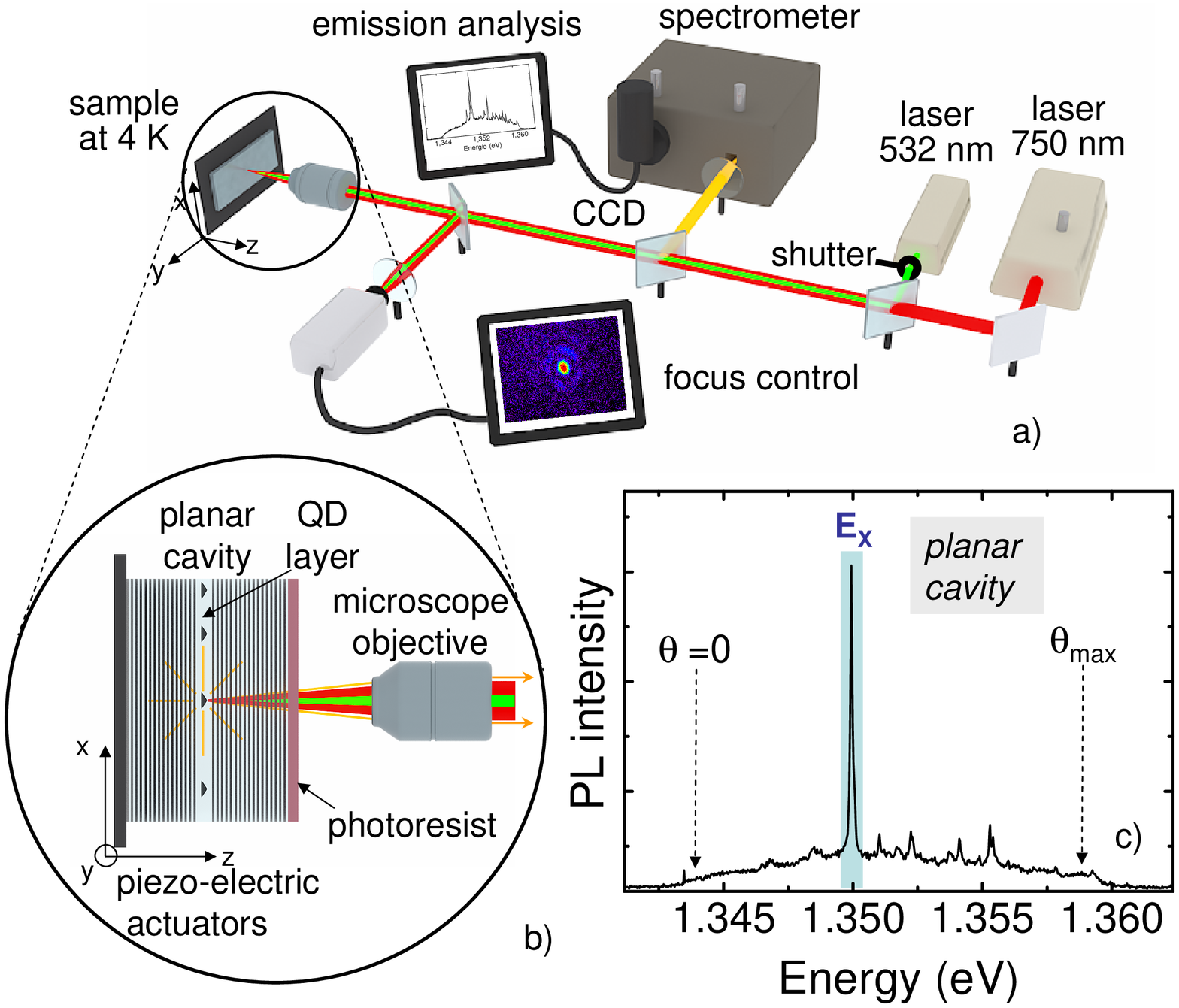, width = 7 cm}}
\end{center}
\caption{\small {a: Schematic diagram of the experimental setup.
b: Schematic illustrating the method. c: PL spectrum measured on
the planar cavity during the photolithography process. }}
\end{figure}

\noindent Earlier CQED investigations with single QDs or
nanocrystals
 were
 based on statistical approaches by which several thousands
of devices were fabricated \cite{11,12,13,crystal1,crystal2}. It
was left to pure statistics and prohibitively long searches to get
the appropriate QD at the appropriate location of the cavity as
the yield was in the low 10$^{-3}$. Lately, the need for control
of the QD-cavity matching has motivated many works
\cite{16,17,18,19,20,pillaretch}. In a pioneer work, Badolato and
coworkers successfully demonstrated deterministic cavity-QD
coupling \cite{10}. The spatial positioning of the QD was achieved
using several technological steps (AFM, electronic lithography)
each  requiring an alignment accuracy of the order of ten
nanometer. The spectral matching was obtained thanks to several
wet-etching steps, preventing the fabrication of more than one
cavity-QD device on a given sample.

In this letter, we show that a far field optical lithography can
be used to position a single QD in a spectrally resonant pillar
microcavity with an accuracy as high as 50 nm. The lithography is
performed in-situ, while measuring the QD emission in a
low-temperature photoluminescence set-up. The nano-positioning is
obtained by using the single QD as a probe of a focused gaussian
laser beam. The cavity-QD spectral matching is achieved in the
same lithography step by selecting a QD of given spectral position
and making use of the non-linearity of the resist exposure
process. The efficiency of the technique is shown by the
observation of large controlled Purcell effect.  The flexibility
of our technique allows  to deterministically couple for the first
time, two QDs to the same optical mode, a milestone for the
implementation of many quantum computing schemes.

\begin{figure}[t]
\begin{center}
\epsfig{file=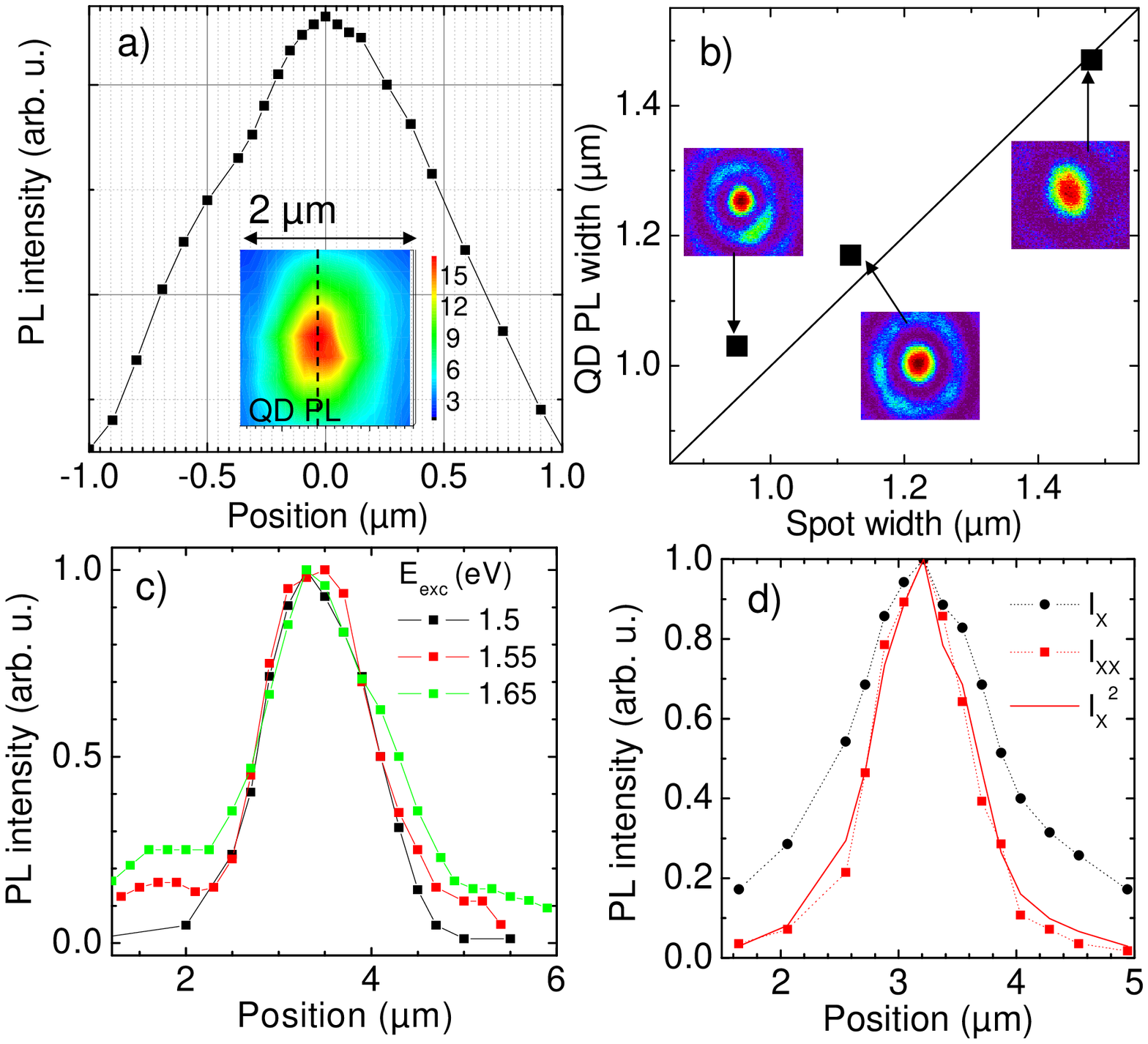, width = 8.5 cm}
\end{center}
\caption{\small {a:  QD PL intensity as a function of the sample
position, along the dotted line on the inset image. Inset: Mapping
of the QD exciton emission intensity as a function of the sample
position. b: QD PL spatial dependence width as a function of the
excitation spot width on the sample. The insets show  6.4$\mu$m x
6$\mu$m images of the excitation laser spot on the control camera.
c: QD PL intensity as a function of the sample position for
various $E_{exc}$. d: red square (resp. black circle): Normalized
PL instensity of the X (resp. XX). Red line : square of the X PL
intensity.}}
\end{figure}
A low density layer of InAs self assembled QDs is grown by
molecular beam epitaxy and located at the center of a
$\lambda$-GaAs microcavity surrounded by two
$Al_{0.1}Ga_{0.9}As/Al_{0.95}Ga_{0.05}As$ Bragg mirrors with
20(24) pairs. The quality factor of the planar microcavity is
Q=4500.  The planar cavity is spin coated with a positive
photoresist and inserted in a helium flow cryostat mounted on a
motorized positioning system with piezoelectric actuators,
providing a 20 nm accuracy. Two laser beams are focused onto the
same area of the sample with a microscope objective. A red laser
beam (750 nm$<\lambda<$830 nm) is used  to measure the QD
emission, without affecting the photoresist, while a green laser
beam ($\lambda$=532 nm) is used to expose the photoresist (Fig.
1). The focusing of the two beams and their superposition is
checked using a monitor camera with a x200 magnification.

\noindent The red excitation gives rise to a photoluminescence
(PL) emission which is spectrally analyzed using a spectrometer
and a charge coupled device camera. The QDs emission is filtered
by the planar cavity mode which energy $E_c$($\theta$) blueshifts
with the collection angle $\theta$ \cite{22}. As a result, QD
emission within a 15 meV spectral range can be observed,
corresponding to the 0.5 numerical aperture of our microscope
objective. The typical PL spectrum in Fig. 1.c. shows a background
emission
 between 1.344 eV and 1.359 eV corresponding to an emission in the cavity mode for
collection angles between $\theta=0^{\circ}$ and
$\theta_{max}=30^{\circ}$. The single QD exciton (X) emission
appears as discrete lines.

 By scanning the sample position in a few micrometer
range, a QD is selected. The sample position is then scanned with
respect to the excitation spot with a 20 nm accuracy. The inset in
Fig. 2.a. shows a mapping of the selected X emission intensity as
a function of the sample position, measured below the X saturation
\cite{23}. Fig. 2.a. shows an intensity profile of the mapping
along the dotted line indicated in the inset. The QD X emission
presents a gaussian like intensity dependence with a width around
$1 \mu m$, reflecting the laser gaussian profile. The same
measurement is performed for various focusing. Fig. 2.b. shows
that the QD PL intensity width is directly given by the size of
the excitation spot. This observation indicates that the capture
into the QD is highly efficient and diffusion outside the
excitation spot negligible. Indeed Fig. 2c shows that the gaussian
dependence
 is unchanged for
excitation energy $E_{exc}$ between $1.5 \ eV$ and $1.6 \ eV$ and
only slightly broadens for $1.65 \ eV$. Scattering outside the
excitation spot only manifests itself by a stronger signal outside
the excitation spot.

Mapping the QD signal allows to determine the QD position by
pointing the maximum of the QD X emission, as shown in Fig. 2.a.
With a zero noise to signal ratio, we would get an accuracy
limited by the QD lateral size. With the noise to signal ratio of
our setup (0.4 $\%$), moving the QD within 50 nm from the center
of the excitation spot results in no measurable change in the QD
signal. As a result, the QD position is pointed with a $\pm$50 nm
 accuracy. This accuracy could be further improved by using
sharper excitation beam profile or by monitoring a non-linear
optical transition of the QD as shown in Fig. 2.e. The biexciton
(XX) spatial dependence (red scatter line), is given by the square
of the exciton spatial dependence (red line) \cite{biexciton}.

\noindent When a QD of
 desired energy is selected and its emission intensity
 maximized to position the QD at the very center of the red spot, the green laser
is switched on to expose the resist. The exposed area will later
be used directly to define the pillar microcavity. This first step
ensures the positioning of the QD at the maximum of the pillar
fundamental mode.

 To
reach the spectral matching, one needs to adjust the pillar radius
R. As shown in Fig. 3.a., the energy of the pillar fundamental
mode increases as the radius decreases \cite{24}. For the example
presented in Fig. 1.c., the energy of the selected QD exciton is
$E_X$=1.350 eV and would be resonant to the fundamental mode of a
pillar with R=1 $\mu$m. The key to achieve spectral matching is to
expose the resist for a well chosen time. Indeed, the chemical
change in the resist depends on the exposure dose in a non-linear
way. Since we use a gaussian laser beam, the size of the exposed
area increases with exposure time or laser power. The radius of
the developed resist area is presented on Fig. 3.b. as a function
of the exposure time for a fixed laser power. Radii ranging from
below 0.5 $\mu$m to 1.5 $\mu$m are reached. Thus, to match the QD
selected in Fig. 1.c., the sample is exposed for 30 s in order to
obtain a 1 $\mu$m radius pillar. After exposing the resist for
several QDs, the photoresist is developed, leaving circular holes
centered on the selected QDs. After a lift off step, these holes
define the mask to fabricate the pillar through chloride reactive
ion etching. A scanning electron microscopy image of pillars of
various radii designed to match QDs of various energies is
presented in Fig. 3.c. Fig. 3.d. shows the obtained mode energy as
a function of the target mode energy. A standard deviation as
small as 0.65 meV is achieved
 limited by the verticality of the pillar
sidewall and the accuracy of the green laser focus. Finally, Fig.
3.e. presents the
 quality factor Q of the pillars  and their Purcell
factor $F_p$. A mean $F_p$ of 6.4 is achieved. The standard
deviation of 1.4 is mainly caused by oscillations of Q predicted
in \cite{lalanne}. Yet, obtaining
 homogeneous $F_p$ is not critical to achieve large scale
integration of efficient single photon sources: the extraction
efficiency
 $F_p/(F_p+1)$
is close to 1 for $F_p \gg 1$.

To prove the high quality of the cavity$-$exciton spatial
matching,  optical measurements are performed on the pillars. The
PL spectrum of a selected QD is shown in Fig. 4.a during the
lithography step (bottom). For this QD, the exposure time was
chosen to obtain a pillar with R=0.85 $\mu$m (pillar 1), to match
the exciton energy of 1.35 eV. The middle and upper parts of Fig.
4.a. present the PL spectra measured at 10 K and 32 K after
etching. At 10 K, the X emission is greatly enhanced and the
fundamental optical mode (M) of the pillar appears as a base to
the X line. Spectral matching is reached at 10 K, as intended.
When increasing temperature up to 32 K, the two lines become
detuned \cite{13} and well resolved. To get a better insight, PL
spectra were acquired over a continuous range of temperatures from
18 K to 32 K (Fig. 4.b.). A strong increase of the QD emission is
observed at resonance with the optical mode, showing the
acceleration of the radiative lifetime for the X\cite{10}. The
linewidth of the optical mode out of resonance is $\gamma_c$=0.89
meV (Q=1500) so that the expected Purcell factor for a QD
perfectly matched to the optical mode is 8.5 \cite{25}. To
estimate the actual F$_P$ experienced by the QD, we follow the
method presented in \cite{26}.
 The maximum X PL signal at saturation is measured as a
function of the X$-$cavity detuning $\Delta$ (Fig. 4c). This
maximum intensity is proportional to $p_X^{sat}(\Delta) \
F_{P,\Delta}$, where $p_X^{sat}(\Delta)$ is the occupation
probability of the QD exciton at saturation and $F_{P,\Delta}$ is
the Purcell factor depending on detuning \cite{26}.
\begin{figure}[t]
\rotatebox{270}{\epsfig{file=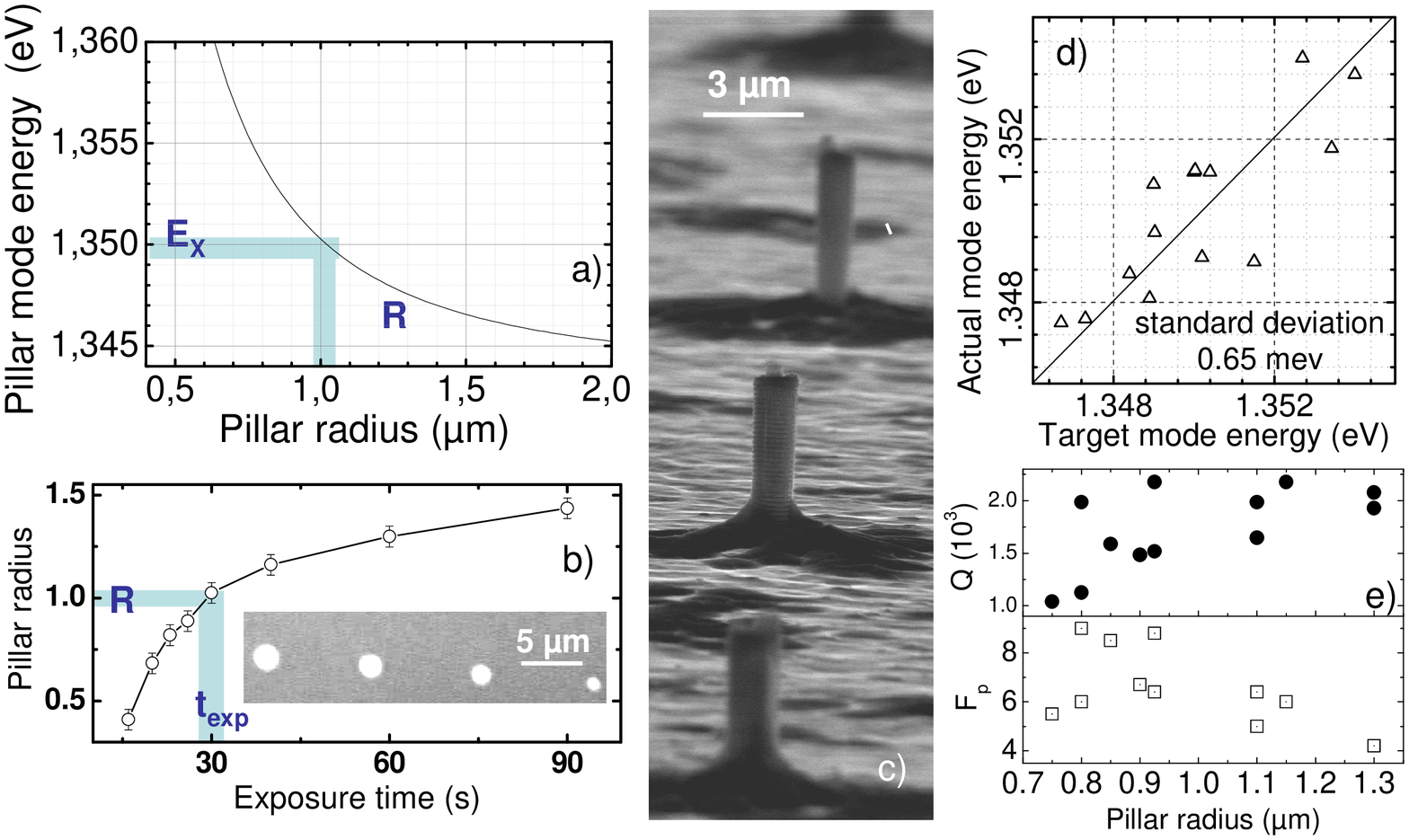, width = 5 cm}}
  \caption{ \small {a: Energy of the pillar
fundamental mode as a function of the pillar radius. b: Pillar
radius as a function of the resist exposure time for a 532 nm laser
power of 500 $\mu$W. insert: Image of the Ni mask obtained after the
lift-off of the resist. c: Scanning electron microscope image of
pillars with various radii designed to match QDs with various
emission energies. d: Actual mode energy as a funtion of target mode
energy. e: Quality factor and Purcell factor as a function of the
pillar radius. }}
\end{figure}
F$_P$ is the maximum Purcell factor for the QD at exact spatial
and spectral resonance with the optical mode. We take
$F_{P,\Delta}=\frac{F_P}{1 + 4 {\frac{\Delta}{\gamma_c}}^2}$
 and $p_X^{sat}(\Delta)=\frac{1}{\sqrt{2+2 F_{P,\Delta}}}$
assuming that the XX decay rate is half the X one. The best fit to
the experimental data gives  F$_P=9\pm 3$, close to the expected
value, showing the high quality of the QD spatial matching to the
optical mode. Indeed, considering an accuracy of 50 nm for the
spatial positioning of the QD, we calculate that the QD is at
least at 98 $\%$ of the maximum of the mode intensity. This value
remains as large as 94 $\%$ for a 0.5 $\mu$m radius pillar .

\begin{figure*}[t]
\rotatebox{270}{\epsfig{file=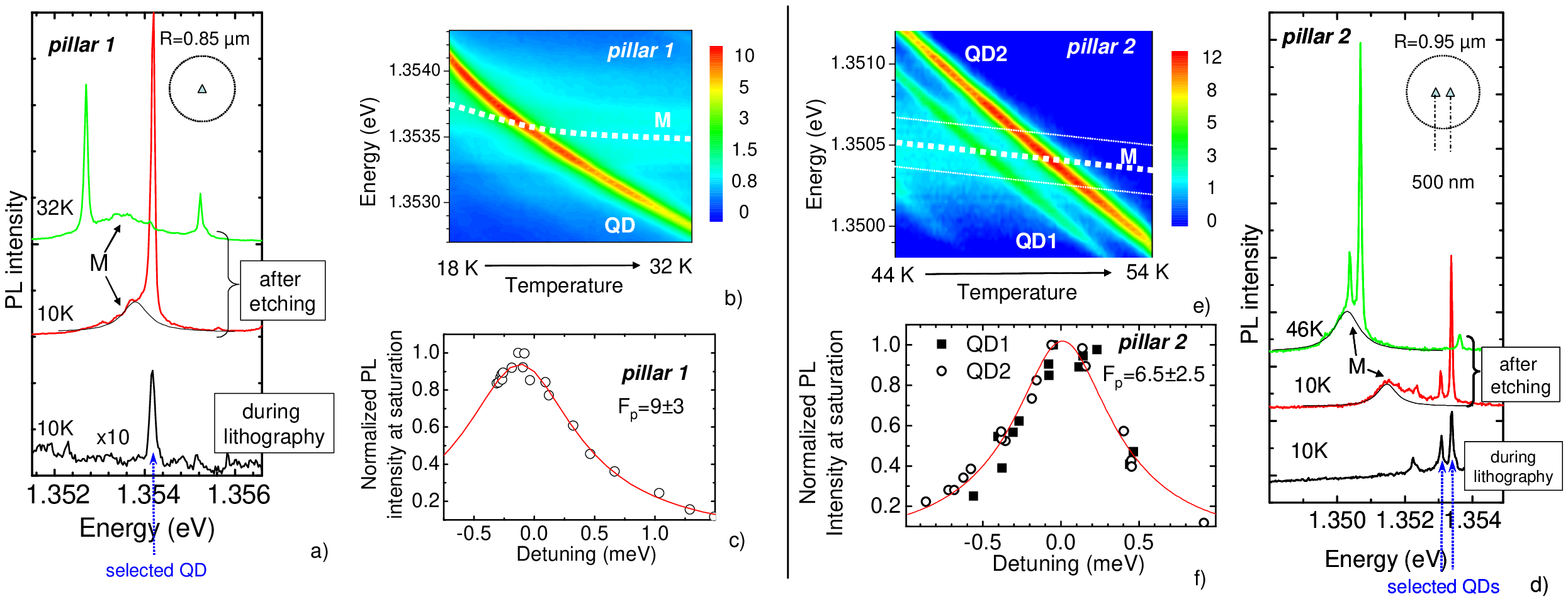, width = 6.6 cm}}
  \caption{ \small {a: (resp d:) Bottom : PL spectrum measured
during the photolithography process at 10 K. Middle and top : PL
spectra of pillar 1 (resp. pillar 2) after etching at 10 K and 32
K (resp. 46 K)(a lorentzian fit to the mode M is also shown). b:
(resp. e:) PL spectra between 18 K and 32 K (resp. 44 K and 54 K)
measured on pillar 1 (resp. pillar 2). c: Pillar 1: Circles : PL
intensity of the X line measured at saturation as a function of
detuning with the optical mode : $\Delta=E_M-E_X$. Line : fit to
the data with $F_P$= 9$\pm$3.  f: Pillar 2: Squares (circles): PL
intensity measured at saturation of the QD1 (QD2) X line as a
function of detuning with the optical mode :
$\Delta_{1(2)}=E_M-E_{QD1(2)}$. Line : fit to the data with
$F_P$=6.5$\pm$2.5 for both QDs.  }}
\end{figure*}

We now demonstrate that our technique can be used to couple two
QDs to the same optical mode in a deterministic way. During the
lithography step, two QDs can be found to be close to each other
both spectrally and spatially. As in the case of a single QD, our
method can be used to define a resonant pillar embedding both QDs:
as the QD positions are identified, the resist is exposed with the
laser spot centered halfway between both QDs. This is what has
been implemented for two QDs found to be located 500 nm apart and
presenting exciton resonances separated by 300 $\mu$eV (Fig. 4d).
Both QDs were inserted in a R=0.95 $\mu$m radius pillar (pillar
2). After etching, the emission of both QDs is slightly detuned
from the pillar mode at 10 K. At 46 K, when bringing the two Xs
close to resonance with the optical mode, an increase of both X
emissions is observed (top spectrum). Fig. 4e. displays the
temperature dependence of the emission spectra measured between 45
K and 57 K which highlight the crossing of both lines with the
optical mode. For both QDs, a strong increase of the signal is
observed at resonance (47 K for QD1 and 49 K for QD2), showing
that both Xs experience a Purcell effect. Fig. 4.f. shows the
normalized signal at saturation for both QDs, as a function of
their detuning with the optical mode \cite{27}. Both QDs present
the same signal variation as a function of detuning, proving that
we have succeeded in equally coupling two QDs to the same optical
mode. The fit to the data gives F$_P=6.5 \pm 2.5$ for both QDs.
The quality factor of this pillar is Q=2350 ($\gamma_c$=0.57 meV)
so that the maximum F$_p$ is supposed to be 9. Since the QDs have
been located 250 nm away from the centre of the pillar, their
overlap to the fundamental mode is reduced to 82$\%$, so that the
expected F$_p$ is 7.3, very close to the measured value.

In conclusion, we have shown that in-situ far field optical
lithography can be used to achieve both the spatial and spectral
matching of a single QD with a pillar microcavity. A dozen devices
have been fabricated on the same sample, within a single
lithography step. Each device evidences a strong Purcell effect,
showing the scalability of the technique. For the present sample,
we intentionally decided to match quantum dot of various energies
to show the flexibility of the spectral tuning. To go toward large
scale integration of efficient single photon sources,
 the present technique could be automatically operated to select QDs within a narrow spectral range
and fabricate arrays of cavity-QD devices at the desired
wavelength. Processing higher quality factor planar cavities
\cite{Q150000} will straightforwardly lead to scalable
deterministic strong coupling regime. This technique could also be
used to register many kinds of nanostructures in an easy, flexible
and scalable way.

\begin{acknowledgments}
\noindent This work is partially supported by the ANR MICADOS and
 the SANDIE European Network. The authors thank P. Voisin and K.
Karrai for fruitful discussions.
\end{acknowledgments}


\begin{thebibliography}{27}
\expandafter\ifx\csname
natexlab\endcsname\relax\def\natexlab#1{#1}\fi
\expandafter\ifx\csname bibnamefont\endcsname\relax
  \def\bibnamefont#1{#1}\fi
\expandafter\ifx\csname bibfnamefont\endcsname\relax
  \def\bibfnamefont#1{#1}\fi
\expandafter\ifx\csname citenamefont\endcsname\relax
  \def\citenamefont#1{#1}\fi
\expandafter\ifx\csname url\endcsname\relax
  \def\url#1{\texttt{#1}}\fi
\expandafter\ifx\csname
urlprefix\endcsname\relax\def\urlprefix{URL }\fi
\providecommand{\bibinfo}[2]{#2}
\providecommand{\eprint}[2][]{\url{#2}}


\bibitem[{\citenamefont{{For a review, see Handbook of Self Assembled Semiconductor Nanostructures for novel devices in photonics and electronics}}(2008)}]{1}
\bibinfo{author}{\bibnamefont{{For a review, see Handbook of Self Assembled Semiconductor Nanostructures for Novel Devices in Photonics and Electronics}}},
  \bibinfo{journal}{Editor M. Henini,  Elsevier},
   (\bibinfo{year}{2008}).




\bibitem[{\citenamefont{{A. Peter Lodahl \it et al}}(2004)}]{colloidal}
\bibinfo{author}{\bibnamefont{{A. Peter Lodahl \it et al}}},
  \bibinfo{journal}{Nature} \textbf{\bibinfo{volume}{430}},
  \bibinfo{pages}{654} (\bibinfo{year}{2004})

\bibitem[{\citenamefont{{B. Lounis  and W. E. Moerner}}(2000)}]{molecule}
\bibinfo{author}{\bibnamefont{{B. Lounis  and W. E. Moerner}}},
  \bibinfo{journal}{Nature} \textbf{\bibinfo{volume}{407}},
  \bibinfo{pages}{491} (\bibinfo{year}{2000}).




\bibitem[{\citenamefont{{P. Michler \it et al}}(2000)}]{2}
\bibinfo{author}{\bibnamefont{{P. Michler \it et al}}},
  \bibinfo{journal}{Science} \textbf{\bibinfo{volume}{290}},
  \bibinfo{pages}{2282} (\bibinfo{year}{2000})





\bibitem[{\citenamefont{{M. Pelton \it et al}}(2002)}]{3}
\bibinfo{author}{\bibnamefont{{M. Pelton \it et al}}},
  \bibinfo{journal}{Phys. Rev. Lett.} \textbf{\bibinfo{volume}{89}},
  \bibinfo{pages}{233602} (\bibinfo{year}{2002})

\bibitem[{\citenamefont{{E. Moreau \it et al}}(2001)}]{4}
\bibinfo{author}{\bibnamefont{{E. Moreau \it et al}}},
  \bibinfo{journal}{Appl. Phys. Lett.} \textbf{\bibinfo{volume}{79}},
  \bibinfo{pages}{2865} (\bibinfo{year}{2001}).


\bibitem[{\citenamefont{{E. M. Purcell, H. C. Torrey and R. V. Pound}}(1946)}]{5}
\bibinfo{author}{\bibnamefont{{E. M. Purcell, H. C. Torrey and R. V. Pound}}},
  \bibinfo{journal}{Phys. Rev.} \textbf{\bibinfo{volume}{69}},
  \bibinfo{pages}{37} (\bibinfo{year}{1946}).


\bibitem[{\citenamefont{{A. Imamoglu \it et al}}(1999)}]{6}
\bibinfo{author}{\bibnamefont{{A. Imamoglu \it et al}}},
  \bibinfo{journal}{Phys. Rev. Lett.} \textbf{\bibinfo{volume}{83}},
  \bibinfo{pages}{4204} (\bibinfo{year}{1999}).


\bibitem[{\citenamefont{{M. Feng, I. D'Amico, P. Zanardi and F. Rossi }}(2003)}]{7}
\bibinfo{author}{\bibnamefont{{M. Feng, I. D'Amico, P. Zanardi and F. Rossi}}},
  \bibinfo{journal}{Phys. Rev. A} \textbf{\bibinfo{volume}{67}},
  \bibinfo{pages}{014306} (\bibinfo{year}{2003}).

\bibitem[{\citenamefont{{M. N. Leuenberger}}(2006)}]{8}
\bibinfo{author}{\bibnamefont{{M. N. Leuenberger}}},
  \bibinfo{journal}{Phys. Rev. B} \textbf{\bibinfo{volume}{73}},
  \bibinfo{pages}{075312} (\bibinfo{year}{2006}).

\bibitem[{\citenamefont{{F. Meier and D. D. Awschalom}}(2004)}]{9}
\bibinfo{author}{\bibnamefont{{F. Meier and D. D. Awschalom}}},
  \bibinfo{journal}{Phys. Rev. B} \textbf{\bibinfo{volume}{70}},
  \bibinfo{pages}{205329} (\bibinfo{year}{2004}).


\bibitem[{\citenamefont{{A. Badolato \it et al}}(2005)}]{10}
\bibinfo{author}{\bibnamefont{{A. Badolato \it et al}}},
  \bibinfo{journal}{Science} \textbf{\bibinfo{volume}{308}},
  \bibinfo{pages}{1158} (\bibinfo{year}{2005}).

\bibitem[{\citenamefont{{T. Yoshie, \it et al}}(2004)}]{11}
\bibinfo{author}{\bibnamefont{{T. Yoshie, \it et al}}},
  \bibinfo{journal}{Nature} \textbf{\bibinfo{volume}{432}},
  \bibinfo{pages}{200} (\bibinfo{year}{2004}).

\bibitem[{\citenamefont{{J. P. Reithmaier, \it et al}}(2004)}]{12}
\bibinfo{author}{\bibnamefont{{J. P. Reithmaier, \it et al}}},
  \bibinfo{journal}{Nature} \textbf{\bibinfo{volume}{432}},
  \bibinfo{pages}{197} (\bibinfo{year}{2004}).

\bibitem[{\citenamefont{{E. Peter, \it et al}}(2005)}]{13}
\bibinfo{author}{\bibnamefont{{E. Peter, \it et al}}},
  \bibinfo{journal}{Phys. Rev. Lett.} \textbf{\bibinfo{volume}{95}},
  \bibinfo{pages}{067401} (\bibinfo{year}{2005}).


\bibitem[{\citenamefont{{N. LeThomas, \it et al}}(2006)}]{crystal1}
\bibinfo{author}{\bibnamefont{{N. LeThomas, \it et al}}},
  \bibinfo{journal}{Nano Lett.} \textbf{\bibinfo{volume}{6}},
  \bibinfo{pages}{557} (\bibinfo{year}{2006}).

\bibitem[{\citenamefont{{Y.-S. Park, A.K. Cook and H. Wang}}(2006)}]{crystal2}
\bibinfo{author}{\bibnamefont{{Y.-S. Park, A.K. Cook and H. Wang}}},
  \bibinfo{journal}{Nano Lett.} \textbf{\bibinfo{volume}{6}},
  \bibinfo{pages}{2075} (\bibinfo{year}{2006}).






\bibitem[{\citenamefont{{A. Hartmann, \it et al}}(1997)}]{16}
\bibinfo{author}{\bibnamefont{{A. Hartmann, \it et al}}},
  \bibinfo{journal}{Appl. Phys. Lett.} \textbf{\bibinfo{volume}{71}},
  \bibinfo{pages}{1314} (\bibinfo{year}{1997}).

\bibitem[{\citenamefont{{S. Kiravittaya \it et al}}(2006)}]{17}
\bibinfo{author}{\bibnamefont{{S. Kiravittaya \it et al}}},
  \bibinfo{journal}{Appl. Phys. Lett.} \textbf{\bibinfo{volume}{89}},
  \bibinfo{pages}{233102} (\bibinfo{year}{2006}).

\bibitem[{\citenamefont{{Z. G. Xie and G. S. Solomon}}(2005)}]{18}
\bibinfo{author}{\bibnamefont{{Z. G. Xie and G. S. Solomon}}},
  \bibinfo{journal}{Appl. Phys. Lett.} \textbf{\bibinfo{volume}{87}},
  \bibinfo{pages}{093106} (\bibinfo{year}{2005}).

\bibitem[{\citenamefont{{A. Rastelli \it et al}}(2007)}]{19}
\bibinfo{author}{\bibnamefont{{A. Rastelli \it et al}}},
  \bibinfo{journal}{Appl. Phys. Lett.} \textbf{\bibinfo{volume}{90}},
  \bibinfo{pages}{073120} (\bibinfo{year}{2007}).


\bibitem[{\citenamefont{{K. Hennessy \it et al}}(2005)}]{20}
\bibinfo{author}{\bibnamefont{{K. Hennessy \it et al}}},
  \bibinfo{journal}{Appl. Phys. Lett.} \textbf{\bibinfo{volume}{87}},
  \bibinfo{pages}{021108} (\bibinfo{year}{2005}).

\bibitem[{\citenamefont{{H. Lohmeyer \it et al}}(2008)}]{pillaretch}
\bibinfo{author}{\bibnamefont{{H. Lohmeyer \it et al}}},
  \bibinfo{journal}{Appl. Phys. Lett.} \textbf{\bibinfo{volume}{92}},
  \bibinfo{pages}{011116} (\bibinfo{year}{2008}).


\bibitem[{\citenamefont{{P. Senellart \it et al}}(2005)}]{biexciton}
\bibinfo{author}{\bibnamefont{{P. Senellart \it et al}}},
  \bibinfo{journal}{Phys. Rev. B} \textbf{\bibinfo{volume}{72}},
  \bibinfo{pages}{115302} (\bibinfo{year}{2005}).




\bibitem[{\citenamefont{{A. Arena \it et al}}(1998)}]{22}
\bibinfo{author}{\bibnamefont{{A. Arena \it et al}}},
  \bibinfo{journal}{Appl. Phys.
Lett.} \textbf{\bibinfo{volume}{72}},
  \bibinfo{pages}{2571} (\bibinfo{year}{1998}).


\bibitem[{\citenamefont{{K. Brunner, G. Abstreiter, G. B\"{o}hm, G. Tr\"{a}nkle and G. Weimann}}(1994)}]{23}
\bibinfo{author}{\bibnamefont{{K. Brunner, G. Abstreiter, G. B\"{o}hm, G. Tr\"{a}nkle and G. Weimann}}},
  \bibinfo{journal}{Phys. Rev.
Lett.} \textbf{\bibinfo{volume}{73}},
  \bibinfo{pages}{1138} (\bibinfo{year}{1994}).

\bibitem[{\citenamefont{{J. M. G\'erard \it et al}}(1996)}]{24}
\bibinfo{author}{\bibnamefont{{J. M. G\'erard \it et al}}},
  \bibinfo{journal}{Appl. Phys. Lett.} \textbf{\bibinfo{volume}{69}},
  \bibinfo{pages}{449} (\bibinfo{year}{1996}).




\bibitem[{\citenamefont{{Ph. Lalanne, J. P.Hugonin and J. M. G\'erard}}(1996)}]{lalanne}
\bibinfo{author}{\bibnamefont{{ Ph. Lalanne, J. P.Hugonin and J. M. G\'erard}}},
  \bibinfo{journal}{Appl. Phys. Lett.} \textbf{\bibinfo{volume}{84}},
  \bibinfo{pages}{4726} (\bibinfo{year}{2004}).


\bibitem[{\citenamefont{{J. M. G\'erard \it et al}}(1998)}]{25}
\bibinfo{author}{\bibnamefont{{J. M. G\'erard \it et al}}},
  \bibinfo{journal}{Phys. Rev. Lett.} \textbf{\bibinfo{volume}{81}},
  \bibinfo{pages}{1110} (\bibinfo{year}{1998}).

\bibitem[{\citenamefont{{C. B\"{o}ckler \it et al}}(2008)}]{26}
\bibinfo{author}{\bibnamefont{{C. B\"{o}ckler \it et al}}},
  \bibinfo{journal}{Appl. Phys. Lett.} \textbf{\bibinfo{volume}{92}},
  \bibinfo{pages}{091107} (\bibinfo{year}{2008}).



\bibitem[{\citenamefont{{ Surprisingly, the PL intensity of the two QDs is
different whereas it was quite similar during the lithography
step. The same effect is observed on another pillar embedding two
QDs  This effect is currently under investigation}}()}]{27}
\bibinfo{author}{\bibnamefont{{ Surprisingly, the PL intensity of the two QDs is
different whereas it was quite similar during the lithography
step. The same effect is observed on another pillar embedding two
QDs  This effect is currently under investigation}}}.



\bibitem[{\citenamefont{{S. Reitzenstein \it et al}}(2007)}]{Q150000}
\bibinfo{author}{\bibnamefont{{S. Reitzenstein \it et al}}},
  \bibinfo{journal}{Appl. Phys. Lett.} \textbf{\bibinfo{volume}{90}},
  \bibinfo{pages}{251109} (\bibinfo{year}{2007}).




\end{thebibliography}
\end{document}